\documentclass{ws-ijmpc}

\usepackage{graphicx,amssymb}
\usepackage{amsmath}
\usepackage{color}

\newcommand{\vrel}{\vec v^\text{rel}}
\newcommand{\vrelt}{\vec v^\text{rel}_t}
\newcommand{\vrelfree}{\vec v^\text{rel,free}}
\newcommand{\n}{\vec n}
\newcommand{\ls}{\lambda \! ^*}
\newcommand{\ps}{P^*_\text{in}}
\newcommand{\tdt}{t+\Delta t}
\newcommand{\Pext}{P_\text{ext}}
\newcommand{\Mv}{M_\text{v}\,}
\newcommand{\Ml}{M_{\lambda}\,}

\begin{document}

\title{Generation of homogeneous granular packings: Contact 
dynamics simulations at constant pressure using fully periodic boundaries}

\author{M.\ Reza Shaebani}

\address{Institute for Advanced Studies in Basic Sciences, Zanjan 45195-1159, Iran}

\address{Department of Theoretical Physics, Budapest University 
of Technology and Economics, H-1111 Budapest, Hungary}

\address{Department of Theoretical Physics, University of 
Duisburg-Essen, 47048 Duisburg, Germany \\
shaebani@comphys.uni-duisburg.de}

\author{Tam\'as Unger}

\address{Department of Theoretical Physics, Budapest University 
of Technology and Economics, H-1111 Budapest, Hungary}

\address{HAS-BME Condensed Matter Research Group, Budapest 
University of Technology and Economics}

\author{J\'anos Kert\'esz}

\address{Department of Theoretical Physics, Budapest University 
of Technology and Economics, H-1111 Budapest, Hungary}

\address{HAS-BME Condensed Matter Research Group, Budapest 
University of Technology and Economics}

\maketitle

\begin{abstract}
The contact dynamics method (CD) is an efficient simulation
technique of dense granular media where unilateral and frictional
contact problems for a large number of rigid bodies have to be
solved. In this paper we present a modified version of the contact
dynamics to generate homogeneous random packings of rigid grains. 
CD simulations are performed at constant external pressure, which allows the
variation of the size of a periodically repeated cell. We follow
the concept of the Andersen dynamics and show how it can be 
applied within the framework of the contact dynamics method. The 
main challenge here is to handle the interparticle interactions 
properly, which are based on constraint forces in CD. We implement 
the proposed algorithm, perform test simulations and investigate 
the properties of the final packings.

\end{abstract}

\ccode{PACS Nos.: 45.70.-n, 45.70.Cc, 02.70.-c, 45.10.-b}

\section{Introduction}
\label{Introduction}

Computer simulation methods have been widely employed in recent
years to study the behavior of granular materials. Among the
numerical techniques, discrete element methods, including
\emph{soft particle molecular dynamics} (MD)
\cite{Cundall79,Silbert02}, \emph{event-driven} (ED)
\cite{Rapaport80,Walton86} and \emph{contact dynamics} (CD)
\cite{Jean92,Moreau94,Jean99,Brendel04}, constitute an important
class where the material is simulated on the level of particles.
In such algorithms the trajectory of each particle is calculated
as a result of interaction with other particles, confining
boundaries and external fields. The differences between the
discrete element methods stem from the way how interactions
between the particles are treated, which leads also to different
ranges of applicability.

In low density granular systems, where interactions are mainly
binary collisions, the event-driven method is an efficient
technique. The particles are modeled as perfectly rigid and the
contact duration is supposed to be zero. The handling of dense
granular systems, where the frequency of collisions is large or
long-lasting contacts appear, becomes problematic in ED
simulations \cite{Haff83,McNamara94}.

In case of dense granular media the approach of soft particle 
molecular dynamics is more favorable and widely used. 
In MD, the time step is usually fixed and the original undeformed
shapes of the particles may overlap during the dynamics. These
overlaps are interpreted as elastic deformations which generate
repulsive restoring forces between the particles. Based on this
interaction, which is defined in the form of a visco-elastic force
law, the stiffness of the particles can be controlled. When the
stiffness is increased MD simulations become slower since the time
step has to be chosen small enough so that the velocities and
positions vary as smooth functions of time.

The contact dynamics method considers the grains as perfectly
rigid. Therefore no overlaps between the particles are expected
and they interact with each other only at contact points. The
contact forces in CD do not stem from visco-elastic force laws but
are calculated in terms of constraint conditions (for more details
see Sec.~\ref{CD-Method}). This method has shown its efficiency in
the simulation of dense frictional systems of hard particles.

Packings of hard particles interacting with repulsive contact
forces are extensively used as models of various complex many-body
systems, e.g. dense granular materials \cite{Mehta94}, glasses
\cite{Zallen83}, liquids \cite{Hansen86} and other random media
\cite{Torquato02}. Jamming in hard-particle packings of granular
materials has been the subject of considerable interest recently 
\cite{Liu98,Combe00}. Furthermore hard-particle packings, and 
especially hard-sphere packings, have inspired mathematicians and 
been the source of numerous challenging theoretical problems 
\cite{Aste00}, from which many are still open.

Real systems in the laboratory and in nature contain far too large
number of particles to model the whole system in computer
simulations. Due to the limited computer capacity the simulations
are often restricted to test a small mesoscopic part of a large
system. Typically, the studies are focused to a local homogeneous
small piece of the material inside the bulk far from the border of
the system. Therefore simulation methods are required that are
able to generate and handle packings of hard particles without
side effects of confining walls.

The usual simulation methods of dense systems involve confining 
boxes where the material is compactified by moving pistons or 
gravity. However, the properties of the material differ in the 
vicinity of walls and corners of the confining cell from those in 
the bulk far from the walls. The application of walls in computer 
simulations leads to inhomogeneous systems due to undesired side 
effects (e.g. layering effect). Moreover, the structure of the 
packings becomes strongly anisotropic in these cases due to the 
orientation of walls and special direction of the compaction. For 
studies where such anisotropy is unwanted other type of compaction 
methods are needed.

In this paper, we present a compaction method where boundary
effects are avoided due to exclusion of side walls. This
simulation method is based on the contact dynamics algorithm where
we applied the concept of the Andersen dynamics \cite{Andersen80}, 
which enables us to produce homogeneous granular packings of hard 
particles with desired internal pressure. The compaction method 
involves variable volume of the simulation cell with periodic 
boundary conditions in all directions.

This paper is organized as follows. First we present some basic
features of CD method in Section~\ref{CD-Method}. Then, 
Section~\ref{Andersen-Method} describes the equations of motion
for a system of particles with variable volume. In 
Section~\ref{CD+Andersen} we present a modified version of CD with
coupling to constant external pressure. In Section~\ref{results} we
report the results of some test simulations.
Section~\ref{conclusions} concludes the paper.

\section{Non smooth contact dynamics}
\label{CD-Method}

\emph{Contact dynamics} (CD), developed by M. Jean and J. J. Moreau
\cite{Moreau94,Jean99,Brendel04}, is a discrete element method in
the sense that the time evolution of the system is treated on the
level of individual particles. Once the total force $\vec F_i$ and
torque $\vec T_i$ acting on the particle $i$ is known, the problem
is reduced to the integration of Newton's equations of motion
which can be solved by numerical methods. Here we use the implicit
first order Euler scheme:
\begin{equation}
  \vec v_i(\tdt)= \vec v_i(t) +
  \frac{1}{m_i} \vec F_i(\tdt)\Delta t
  \label{veloc-update}
\end{equation}
\begin{equation}
  \vec r_i(\tdt) = \vec r_i(t) +
  \vec v_i(\tdt)\Delta t \, ,
  \label{pos-update}
\end{equation}
which gives the change in the position $\vec r_i$ and velocity
$\vec v_i$ of the center of mass of the particle with mass $m_i$
after the time step $\Delta t$. $\Delta t$ is chosen so that the 
relative displacement of adjacent particles during one time step 
is small compared to the particle size and to the radius of 
curvature of the contacting surfaces. Corresponding equations are 
used also for the rotational degrees of freedom, describing the 
time evolution of the orientation and the angular velocity $\vec
\omega_i$ caused by the new total torque $\vec T_i(\tdt)$ acting
on the particle $i$.

The interesting part of the CD method is how the interaction
between the particles are handled. For simplicity we assume that
the particles are noncohesive and dry, we exclude electrostatic
and magnetic forces between them and consider only interactions
via contact forces. The particles are regarded as \emph{perfectly
rigid} and the contact forces are calculated in terms of
constraint conditions. Such constraints are the impenetrability and
the no-slip condition, i.e. the contact force has to prevent the
overlapping of the particles and the sliding of the contact
surfaces. This latter condition is valid only below the Coulomb
limit of static friction, which states that the tangential
component of a contact force $\vec R$ can not exceed the normal
component times the friction coefficient $\mu$:
\begin{equation}
  \left| \vec R_t \right| \le \mu R_n \, .
  \label{Coulomb-cone}
\end{equation}
If the friction is not strong enough to ensure the no-slip
condition the contact will be sliding and the tangential component
of the contact force is given by the expression
\begin{equation}
  \vec R_t=-\mu R_n
  \frac{\vrelt}{\left|\vrelt \right|} \, ,
  \label{tangential-force}
\end{equation}
where $\vrelt$ stands for the tangential component of the relative
velocity between the contacting surfaces. In the CD method the
constraint conditions are imposed on the new configuration at time
$\tdt$, i.e., the unknown contact forces are calculated in a way
that the constraints conditions are fulfilled in the new
configuration \cite{Brendel04}. This is the reason why an implicit
time stepping is used.

In order to let the system evolve one step from time $t$ to $\tdt$
one has to determine the total force and torque acting on each
particle which may consist of external forces (like gravity) and
contact forces from neighboring particles. Let us suppose that all
the unknown contact forces are already determined except for one
force between a pair of particles already in contact or with a
small gap between them. Here we explain briefly how the constraint
conditions help to determine the interaction between these two
particles. A detailed description of the method can be found
in~\cite{Brendel04}.

\begin{figure}
\begin{center}
\includegraphics*[scale=0.35]{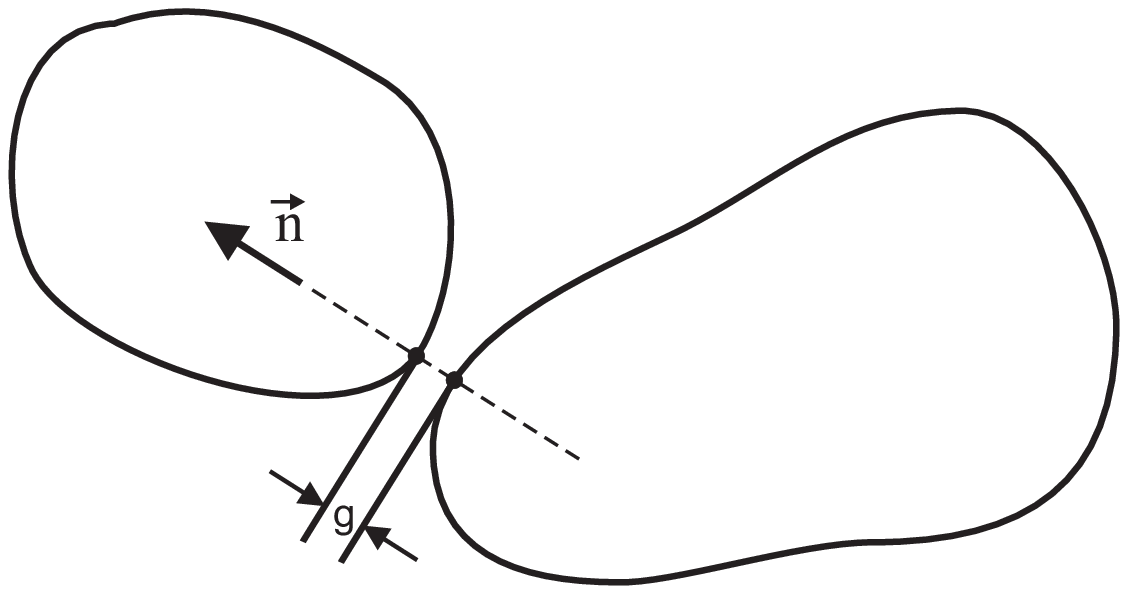}
\end{center}
\caption{Two rigid particles before a possible contact.}
\label{Fig-SchematicContact}
\end{figure}

The algorithm starts with the assumption that the contact force we
are searching for is zero and checks whether this leads to an
overlap of the undeformed shapes of the two particles after one
time step $\Delta t$. This is done based on the time stepping
[Eq.~(\ref{veloc-update})]: The external forces and other contact
forces provide $F_i(\tdt)$ and $T_i(\tdt)$ for both particles thus
the new relative velocity of the contacting surfaces $\vrelfree$ 
can be calculated. Here we use the term \emph{contacting surfaces} 
for simplicity thought the two particles are not necessarily in 
contact. There might be a positive gap $g$ between them, which is 
the length of the shortest line connecting the surfaces of the two 
particles (Fig.~\ref{Fig-SchematicContact}). We will refer to the 
relative velocity of the endpoints of the line as the relative 
velocity of the contact and denote the direction of the line by 
the unit vector $\n$. In the limit of a real contact $g$ is zero 
and $\n$ becomes the contact normal. Negative gap has the meaning 
of an overlap. The superscript \emph{free} in $\vrelfree$ denotes 
that the relative velocity has been calculated assuming no 
interaction between the two particles. We use the sign convention 
that negative normal velocity ($\n \cdot \vrel < 0$) means 
approaching particles.

The new value of the gap (after one time step) is estimated by the
algorithm based on the current gap $g$ and the new relative
velocity $\vrelfree$ according to the implicit time stepping. If
the new gap is positive:
\begin{equation}
  g + \vrelfree \cdot \n \Delta t > 0
  \label{positive-gap}
\end{equation}
then the zero contact force (no interaction) is accepted because
no contact is formed between the two particles. However, if the
estimated new gap is negative then a contact force has to be
applied in order to avoid the violation of the constraint
conditions. Generally, one expects the following relation between
the unknown new contact force $\vec R$ and the unknown new
relative velocity $\vrel$:
\begin{equation}
  \vec R=\frac{-1}{\Delta t} {\bf M}(\vrelfree-\vrel),
  \label{general-contact-force}
\end{equation}
where $\bf M$ is the mass matrix that describes the inertia of the
contact, i.e. ${\bf M}^{-1}\vec R$ is the relative acceleration of
the contacting surfaces due to the contact force $\vec R$. The
mass matrix $\bf M$ depends on the shape, mass and moment of
inertia of the two particles. On one hand, the interpenetration of
the two rigid particles has to be avoided, which gives the
following constraint for the normal component of $\vrel$:
\begin{equation}
  g + \vrel \cdot \n \Delta t = 0.
  \label{gap-close}
\end{equation}
On the other hand, the tangential component of $\vrel$ has to be
zero in order to ensure the no-slip condition
\begin{equation}
  \vrelt = 0.
  \label{no-slip}
\end{equation}
The required contact force that fulfills Eqs.~(\ref{gap-close})
and~(\ref{no-slip}) then reads
\begin{equation}
  \vec R=\frac{-1}{\Delta t} {\bf M}
  (\frac{g}{\Delta t}\n+\vrelfree).
  \label{contact-force}
\end{equation}
This contact force is acceptable only if it fulfills the Coulomb
condition [Eq.~(\ref{Coulomb-cone})]. Otherwise we can not exploit
the non-slip contact assumption. In this case, $\vrelt$ is not
zero, the contact slides and the contact force has to be
recalculated. Eqs.~(\ref{general-contact-force})
and~(\ref{gap-close}) then provide
\begin{equation}
  \vec R=\frac{-1}{\Delta t} {\bf M}
  (\frac{g}{\Delta t}\vec n-\vrelt
  +\vrelfree),
  \label{contact-force-2}
\end{equation}
where the number of unknowns (components of $\vec R$ and $\vrelt$)
exceeds the number of equations. In order to determine the contact
force $\vec R$ one has to solve Eq.~(\ref{contact-force-2})
together with Eq.~(\ref{tangential-force}).

It is recommended to use $g^\text{pos}=max(g,0)$ instead of $g$ in
Eqs.~(\ref{contact-force}) and~(\ref{contact-force-2}). The gap size 
$g$ should always be non-negative and using $g^\text{pos}$
apparently makes no difference \cite{Jean99}. However, due to the
inaccuracy of the calculations small overlaps can be created
between neighboring particles. If $g$ instead of $g^\text{pos}$ is
used then these overlaps are eliminated in the next time step by
imposing larger repulsive contact forces to satisfy
Eq.~(\ref{gap-close}), which pumps kinetic energy into the system.
Using $g^\text{pos}$ instead of $g$ eliminates this artifact on the 
cost that an already existing overlap is not removed (which then 
serves to check the inaccuracies of the simulation \cite{Unger02}), 
only its further growth is prevented. Regarding the above mentioned 
points, we rewrite the equations~(\ref{contact-force}) 
and~(\ref{contact-force-2}) as
\begin{equation}
  \vec R=\frac{-1}{\Delta t}{\bf M}
  (\frac{g^\text{pos}}{\Delta t}\n + \vrelfree) \text{\;\;\; and}
  \label{contact-force-gpos}
\end{equation}

\begin{equation}
  \vec R=\frac{-1}{\Delta t}{\bf M}
  (\frac{g^\text{pos}}{\Delta t}\n - \vrelt + \vrelfree).
  \label{contact-force-2-gpos}
\end{equation}

\begin{figure}
\begin{center}
\includegraphics*[scale=0.38,angle=-90]{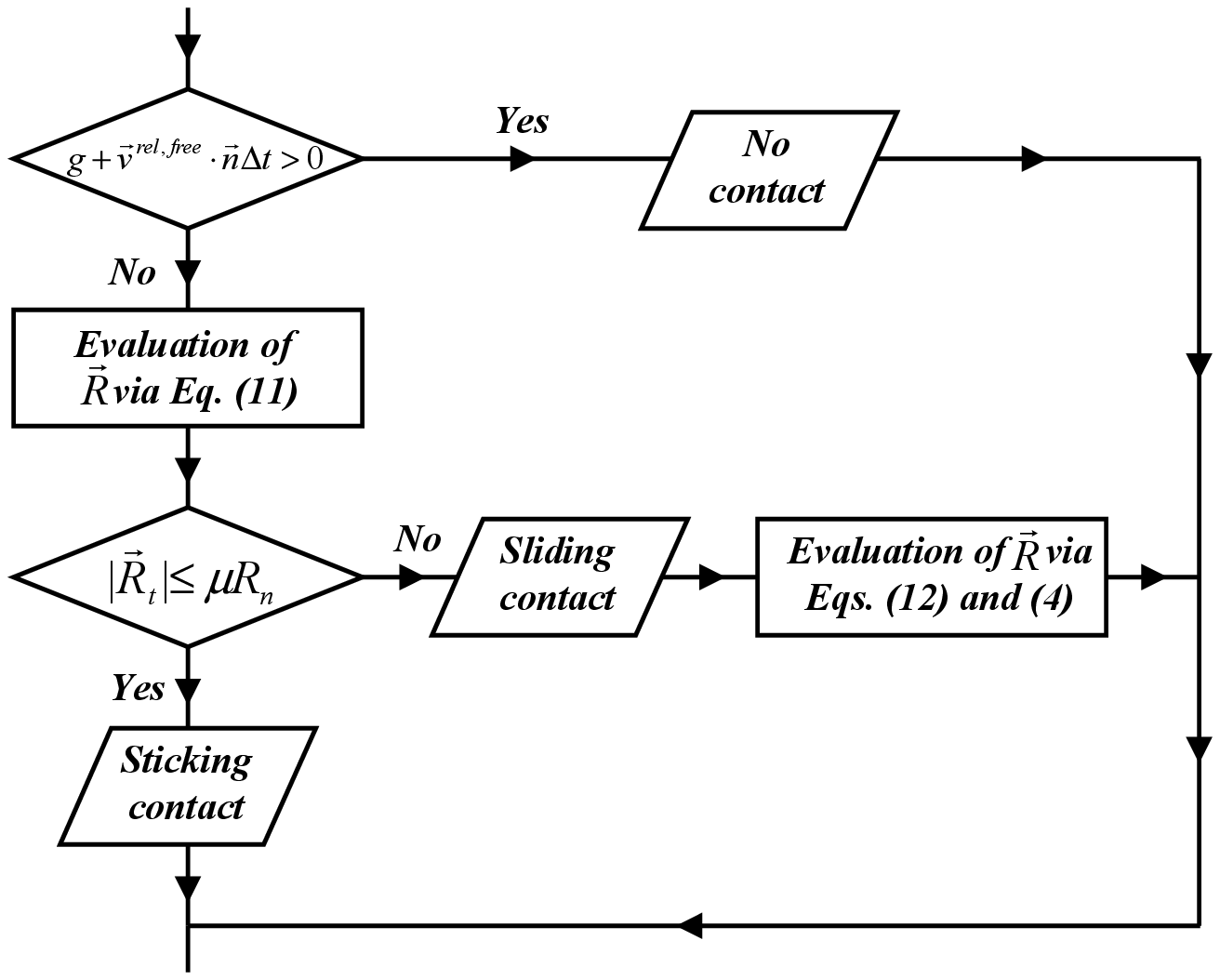}
\end{center}
\caption{The flowchart of the force calculation of a single
contact in contact dynamics.}
\label{Fig-Flowchart}
\end{figure}

A flowchart of the single contact force calculation is given in
Fig.~\ref{Fig-Flowchart}. So far we have explained only how the CD
algorithm determines a single existing or incipient contact, based
on the assumption that all the surrounding contact forces are
known. However, in a dense granular media, many particles contact
simultaneously and form a contact network. In this case, a contact
force cannot be evaluated locally, since it depends on the
adjacent contact forces which are also unknown. To find a globally
consistent force network at each time step, an \emph{iterative
scheme} is applied in CD.

At each iteration step, all contacts are chosen one by one 
and the force at the contact is updated according to the scheme 
shown in Fig.~\ref{Fig-Flowchart}. The update is sequential, i.e., 
the freshly updated contact force is stored immediately as the 
current force and then a new contact is chosen for the next 
update.

After one iteration step, constraint conditions are not
necessarily fulfilled for each contact. In order to find a global
solution the iteration process has to be repeated several times
until the resulting force network converges. The convergence of
the iteration process is smooth, i.e., the precision of the
solution increases with the number of iterations $N_{I}$. Higher
$N_{I}$ provides more precise solution but also requires more
computational effort. The CD method can be used with a constant
number of iterations in subsequent time steps
\cite{Unger02,Brendel04} or with a convergence criterion that
prescribes a given precision to the force calculation
\cite{Moreau94,Jean99,Brendel04}. In this latter case, the number
of iterations $N_{I}$ varies from time step to time step.

After the new force network is determined with a prescribed
precision, the system evolves at the end of the time step 
according to the time-stepping scheme described at the beginning 
of this section. It is important to note that choosing small 
$N_{I}$ and/or large time step causes systematic errors of the 
force calculation which lead to a spurious soft particle behavior
\cite{Unger02} in spite of the original assumption of perfect
rigidity.

To conclude this section, we present briefly the scheme of the
solver.
\begin{displaymath}
\hspace{-3cm}\left[
 \begin{array}{r}
  \hbox{$t:=\tdt$ (time step)\hspace{6.8cm}} \\
  \hbox{Evaluating the gap $g$ for all contacts\hspace{4.7cm}} \\
  \left[
   \begin{array}{r}
    \hbox{$N_I:=N_I+1$ (iteration)\hspace{6.2cm}} \\
    \left[
     \begin{array}{r}
      \hbox{$k:=k+1$ (contact index)\hspace{5.55cm}} \\
      \hbox{Evaluating $\vrelfree_k$ then $\vec R_k$
      (according to the flowchart in
      Fig.~\ref{Fig-Flowchart})\hspace{-0.4cm}} \\
     \end{array}
    \right.
    \\
    \hbox{Convergence test for contact forces\hspace{4.55cm}} \\
   \end{array}
  \right.
  \\
  \hbox{Time-stepping for velocities and positions of all
  particles (using Eqs.~(\ref{veloc-update})
  and~(\ref{pos-update}))\hspace{-2.2cm}} \\
 \end{array}
\right.
\end{displaymath}

\section{The equations of motion at constant external pressure}
\label{Andersen-Method}

In the simulation of granular materials, it is often desirable to
investigate systems which are not surrounded by walls and to apply
periodic boundary conditions in all directions. It is a nice
feature of periodic boundary conditions that they make points of
the space equivalent, the boundary effects are eliminated. That 
way the bulk properties of the material can be studied more easily. 
However, the application of an external pressure becomes 
problematic since the total volume is fixed and the system cannot 
be compressed by pistons or moving walls.

In order to overcome this problem, but at the same time keep the
advantageous periodic boundaries, Andersen \cite{Andersen80}
proposed a method for molecular dynamics simulations. Here, we 
recall his method briefly as its basic ideas will be used later on 
in this paper.

According to the Andersen method the boundaries are still
periodically connected in all directions and no walls are present,
but the volume of the system is a dynamical variable which evolves
in time driven by constant external pressure. When a
system of $N$ atoms is compressed or expanded it is done in an
isotropic and homogeneous way: The distances between the atoms are
rescaled by the same factor regardless of the relative or absolute
positions.

Let us give the equations of motion of a system with particle 
positions $\vec r_1, \vec r_2,..., \vec r_N$ in a D-dimensional 
cubic volume $V$ ($D=2,3$ and each component of $\vec r_i$ is 
between $0$ and $V^{\text{1/D}}$):
\begin{equation}
  \frac{d \vec r_i(t)}{dt} = \frac{\vec p_i(t)}{m_i} +
  \frac{1}{D} {\vec r_i(t)} \frac{d\ln V(t)}{dt},
  \label{equation-motion-1}
\end{equation}
\begin{equation}
  \frac{d \vec p_i(t)}{dt} = \vec F_i(t)- \frac{1}{D}
  {\vec p_i(t)} \frac{d\ln V(t)}{dt},
  \label{equation-motion-2}
\end{equation}
\begin{equation}
  \frac{\Mv d^2 V(t)}{dt^2} = P_\text{in}(t) - \Pext = 
  \Delta P(t).
  \label{equation-motion-3}
\end{equation}
Eq.~(\ref{equation-motion-1}) describes the change in the position
$\vec r_i$. The first term on the right is the usual one, the
momentum $\vec p_i$ divided by the mass of the $i$th particle. The
last term is the extension by the Andersen method that rescales
the position according to the relative volume change.

Eq.~(\ref{equation-motion-2}) provides the time evolution of the
momentum due to two terms. The first one is the usual total force
$\vec F_i$ acting on the $i$th particle which originates from the
interaction with other particles and/or from external fields. The
additional term leads to further acceleration of the particle if
the volume is changing. E.g., if the system is compressed the
kinetic energy of the particles is increased due to the work done
by the compression. The energy input is achieved by rescaling all
particle momenta regardless of their positions. This is in
contrast to usual pistons where the energy enters at the boundary.

Eq.~(\ref{equation-motion-3}) can be interpreted as Newton's
second law that governs the change of the volume. It describes the
time evolution of an imaginary piston which has the inertia 
parameter $\Mv$ and is driven by the generalized force $\Delta P(t)=
P_\text{in}(t) - \Pext$. This latter is the pressure difference 
between the constant external pressure $\Pext$ 
and the internal pressure of the system $P_\text{in}(t)$. The
pressure difference $\Delta P(t)$ drives the system towards the
external pressure, the sensitivity of the system to this driving
force is controlled by the inertia parameter $\Mv$.

In the limit of infinite inertia $\Mv \rightarrow \infty$ together
with the initial condition $dV(t_0)/dt=0$ the volume of the system
remains constant and Eqs.~(\ref{equation-motion-1})
and~(\ref{equation-motion-2}) correspond to the usual Newtonian
dynamics of the particles.

In order to get more insight into the Andersen dynamics let us
consider a simple example of a system of non-interacting particles
with all $\vec F_i(t)=0$. Initially, the velocities and the volume
velocity $dV(t_0)/dt$ are set to zero. Because the internal
pressure $P_\text{in}$ is zero the system with finite inertia $\Mv$
and under external pressure $\Pext>0$ will start contracting
according to Eq.~(\ref{equation-motion-3}). The acceleration of
the particles [Eq.~(\ref{equation-motion-2})] remains zero during
the time evolution; One might say that the particles are standing
there all the time. However, the distances between them are
decreasing because of the contraction of the ``world'' around
them. This is caused by the second term on the right hand side of
Eq.~(\ref{equation-motion-1}) while the first term remains zero.

This suggests the picture of an imaginary background membrane that
contracts or dilates homogeneously together with the volume and
carries the particles along. The velocity of this background at
position $\vec r_i$ is given by $\lambda(t) \vec r_i$ where
$\lambda$ is the dilation rate defined by the rate of the relative
change in the system size $L$:
\begin{equation}
  \lambda(t) \equiv \frac{\dot L(t)}{L(t)} =
  \frac{1}{D} \frac{d\ln V(t)}{dt}
  \label{dilation-rate}
\end{equation}
and $\text{D}$ is the dimension of the system. Then the right hand
side of Eq.~(\ref{equation-motion-1}) can be interpreted as the
sum of two velocities: the second one is the velocity of the
background at the position of the particle and the first one is
the intrinsic velocity of the particle measured compared to the
background. The sum of these two forces gives the changing rate of
the absolute position $\vec r_i$. In the rest of the paper the
velocity $\vec v_i$ will refer always to the intrinsic velocity.
We rewrite Eq.~(\ref{equation-motion-1}) in the following form
\begin{equation}
  \frac{d \vec r_i (t)}{dt} = \vec v_i (t) +
  \lambda(t) {\vec r_i (t)}.
  \label{equation-motion-new-1}
\end{equation}

Next we turn to the modelling of granular systems. Our goal is to
achieve static granular packings that are compressed from a loose
gas-like state. Here again it is advantageous to exclude confining
walls and in order to apply pressure and achieve contraction of
the volume we will use the concept of the Andersen method.
However, the equations of motion will be slightly changed in order
to make them suit better to our goals.

In granular materials the interactions between the particles are
dissipative. When the material is poured into a container or is
compressed by a piston the particles gain kinetic energy due to
the work done by gravity or the piston. All this energy has to be
dissipated (turned into heat) by the interactions between the 
particles before the material can settle into a static dense 
packing of the particles. This relaxation process requires a 
massive computational effort when large packings are modeled in 
computer simulations. One encounters the same problem if the 
Andersen dynamics is applied straight to granular systems. 
The role of the second term on the right hand side of 
Eq.(14) is to conserve the total energy of the system 
by taking into account the energy gain of the particles due 
to compaction. The relaxation time can be reduced if this 
term is omitted, because then the total amount of energy pumped 
into the system is reduced. In this case the particles are 
accelerated only by the forces $\vec F_i$ but they receive 
no additional energy due to the decreasing volume. 
Thus the following equation will be applied for granular
compaction here
\begin{equation}
  \frac{d \vec v_i (t)}{dt} = \frac{1}{m_i} \vec F_i (t)
  \label{equation-motion-new-2}
\end{equation}
which results in a more effective relaxation rather than
Eq.~(\ref{equation-motion-2}). This change is advantageous also 
from the point of view of momentum conservation. If the system is 
compactified by using Eq.~(\ref{equation-motion-2}) from an 
initial condition where the total momentum of the particles is 
non-zero, then this momentum will be increased inverse 
proportionally to the size of the system $L$. Thus the total 
momentum e.g. due to initial random fluctuations is magnified 
which can lead to non-negligible overall rigid body motion of 
the final static packing. This is in contrast to 
Eq.~(\ref{equation-motion-new-2}) which provides momentum 
conservation in the absence of external fields.

We note that neglecting the term in 
Eq.~(\ref{equation-motion-2}) leads to an artificial dynamics in the sense
that the energy corresponding to the work of the compaction is not
delivered to the particles. However, our main goal here is to produce a
static configuration of grains and contact forces that can be used for
further studies, thus we are not interested in that part of the dynamics, where 
compaction rate is significant and the neglected term makes a 
difference.

Concerning the equation that describes the time evolution of the
system size we find it more convenient to control $\lambda$
instead of $dV/dt$. This is actually not an important change and
leads to very similar dynamics. Our third equation reads
\begin{equation}
  \Ml \frac{d\lambda(t)}{dt} = \Delta P(t).
  \label{equation-motion-new-3}
\end{equation}

The equations of motion~(\ref{equation-motion-new-1}),
(\ref{equation-motion-new-2}) and (\ref{equation-motion-new-3})
describe an effective compaction dynamics for granular systems,
they are able to provide static packings under the desired
pressure $\Pext$ and if they are restricted to the limit of $\Ml
\rightarrow \infty$ we receive back the classical Newtonian
dynamics. 

We note that the force scale in our systems of rigid particles is 
determined by the the external pressure as there is no intrinsic 
force scale. Taking larger value of $\Pext$ leads in principle to 
the same compaction dynamics (apart from rescaling time, velocities 
and forces).  Consequently, the same final packing-configuration is 
expected, independently of the external pressure. Of course, the value 
of $\Pext$ does matter if an intrinsic force scale is present, e.g. 
when cohesion between the particles is incorporated. In such cases the 
final packing will strongly depend on the applied external pressure.


In order to close the equations we need to define
interactions between the particles. The interparticle forces
provide $\vec F_i$ in Eq.~(\ref{equation-motion-new-2}) and they
are also needed to evaluate the inner pressure $P_\text{in}$.

The stress tensor $\sigma_{\alpha\beta}$ is not a priori 
spherical in granular materials. The average 
$\sigma_{\alpha\beta}$ of the system is determined by the 
interparticle forces \cite{Christoffersen81} and the particle 
velocities as
\begin{equation}
  \sigma_{\alpha\beta} = \frac{1}{V}
  (\displaystyle\sum_{k=1}^{N_c}
  F_{k,\alpha} l_{k,\beta} + \displaystyle\sum_{i=1}^{N} m_i v_{i,\alpha}
  v_{i,\beta}),
  \label{stress-tensor}
\end{equation}
where $N$ and $N_c$ denote the number of particles and the number
of contacts, respectively. If two contacting particles at contact 
$k$ are labelled by $1$ and $2$, then $\vec F_k$ is the force 
exerted on particle $2$ by particle $1$, and the vector 
$\vec l_k$ is pointing from the center of mass of particle $1$ to 
that of particle $2$ where periodic boundary conditions and nearest 
image neighbors are taken into account. Thus $l_k$ is the minimum
distance between particles $1$ and $2$:
\begin{equation}
  l_k = |\vec l_k|= min |{\vec r_{2}} -
  {\vec r_{1}} + V^\text{1/3} {\vec a}|,
  \label{min-distance}
\end{equation}
where $\vec a$ is an integer-component translation vector. The
inner pressure is then given by the trace of the stress tensor
divided by the dimension of the system
\begin{equation}
  P_\text{in}= \frac{1}{\text{DV}}[\displaystyle\sum_{k=1}^{N_c}
  \vec F_k \cdot \vec l_k + \displaystyle\sum_{i=1}^{N} m_i \vec
  v_i \cdot \vec v_i],
  \label{inner-pressure}
\end{equation}
which has the meaning of an average normal stress.

The implementation of the above method in computer simulations is
straightforward if the interparticle forces are functions of the
positions and velocities of the particles, e.g., in soft particle
MD simulations. The implementation is less
trivial for the case of the contact dynamics method where
interparticle forces are constraint forces. We devote the next
section to this problem.

\section{Contact dynamics with coupling to a constant external pressure}
\label{CD+Andersen}

In this section we present a modified version of the contact dynamics 
algorithm which enables us to perform CD simulations at constant external 
pressure. According to Sec.~\ref{Andersen-Method} let us suppose that the 
system is subjected to a constant external pressure $\Pext$ and its time 
evolution is given by Eqs.~(\ref{equation-motion-new-1}),
(\ref{equation-motion-new-2}) and (\ref{equation-motion-new-3}).

Here we will follow the description of the CD method given in
Sec.~\ref{CD-Method} and discuss the required modifications. Once
the force calculation process is completed, the implicit Euler
integration can proceed one time step further. Now the time
stepping has to involve also the equations of motion of the system
size. By discretizing the Eqs.~(\ref{dilation-rate})
and~(\ref{equation-motion-new-3}) in the same implicit manner as
for the particles [Eqs.~(\ref{veloc-update})
and~(\ref{pos-update})] we obtain the new values of the system
size $L$ and the dilation rate $\lambda$:
\begin{equation}
 \lambda(\tdt)= \lambda(t) +
 \frac{\Delta P(\tdt)}{\Ml} \Delta t,
  \label{dilation-rate-update}
\end{equation}
\begin{equation}
 L(\tdt)= L(t)[1+\lambda(\tdt) \Delta t]
  \label{system-update}
\end{equation}
where the ``velocity'' $\lambda(t)$ and the ``position'' $L(t)$
are updated by the new ``force'' $\Delta P(\tdt)$ and by the new
``velocity'' $\lambda(\tdt)$, respectively.

The discretized equations governing the translational degrees of
freedom of the particles [Eqs.~(\ref{veloc-update})
and~(\ref{pos-update})] are rewritten according to
Eqs.~(\ref{equation-motion-new-1})
and~(\ref{equation-motion-new-2}) in the following form:
\begin{equation}
 \vec v_i(\tdt)= \vec v_i(t) +
 \frac{1}{m_i} \vec F_i(\tdt) \Delta t \text{\;\;\; and}
  \label{particles-vel-update}
\end{equation}
\begin{equation}
 \vec r_i(\tdt)= \vec r_i(t)[1+\lambda(\tdt) \Delta t]
 +  \vec v_i(\tdt) \Delta t.
  \label{particles-pos-update}
\end{equation}
The time stepping for the rotational degrees of freedom remains
unchanged because the dilation (contraction) of the system has no
direct effect on the rotation of the particles.

In the CD method, as we explained in Sec.~\ref{CD-Method} the
particles are perfectly rigid and are interacting with constraint
forces, i.e. those forces are chosen between contacting particles
that are needed to fulfill the constraint conditions. E.g. the
contact force has to prevent the interpenetration of the
contacting surfaces.

If a constant external pressure is used then the calculation of the 
constraint forces has to be reconsidered because the relative
velocity of the contacting surfaces is influenced by the variable
volume. When the system is dilating or contracting, particles gain
additional relative velocities compared to each other. For a pair
of particles, this velocity is $\lambda \vec l$ where $\vec l$ is
the vector connecting the two centers of mass. The same change
appears in the relative velocity of the contacting surfaces as the 
size of the particles is kept fixed. If this change led to 
interpenetration then it has to be compensated by a larger contact 
force. It may also happen that existing contacts open up due to 
expansion of the system resulting in zero interaction force for 
those pair of particles.

In the calculation of a single contact force, the relative
velocity $\lambda \vec l$ (i.e. the contribution of the changing
system size) has to be added to $\vrelfree$.  The new relative 
velocity of the contact assuming no interaction between the two 
particles $\vrelfree$ is calculated here in the same way as
in Sec.~\ref{CD-Method}, i.e., based on the intrinsic velocities 
of the particles. Thus the effect of the dilation/contraction of 
the system is not taken into account in $\vrelfree$. Therefore one 
has to replace $\vrelfree$ with ($\vrelfree + \lambda \vec l$) in 
all equations of Sec.~\ref{CD-Method} in order to impose the 
constraint conditions properly. Let us first suppose that 
the system has infinite inertia ($\Ml = \infty$) thus the dilation 
rate $\lambda$ is constant.  In this case the modified equations 
of the force update (containing already the term $\vrelfree + 
\lambda \vec l$) provide the right constraint forces at the end of 
the iteration process. These forces will alter the relative  
velocity ($\vrelfree + \lambda \vec l$) in such a way that the 
prescribed constrain conditions will be fulfilled in the new 
configuration at $\tdt$.

More consideration is needed if finite inertia $\Ml$ is used and the 
dilation rate $\lambda$ is time-dependent. The problem is that in 
order to calculate the proper contact force one has to know the 
new dilation rate. The new dilation rate, however, depends on the 
new value of the inner pressure [Eq.~(\ref{dilation-rate-update})] 
which, in turn, depends on the new value of the contact forces. 
This problem can be solved by incorporating $\lambda$ and 
$P_\text{in}$ into the iteration process. Instead of using the old 
values $\lambda(t)$ and $P_\text{in}(t)$ during the iteration, we 
always use the expected values $\ls$ and $\ps$. These represent 
our best guess for the new dilation rate $\lambda(\tdt)$ and for 
the new inner pressure $P_\text{in}(\tdt)$. $\ps$ is defined based 
on the current values of the contact forces $\vec F^*_k$ during 
the force iteration. Whenever a contact force is updated we 
recalculate the expected inner pressure $\ps$. With the help of 
the current contact forces $\vec F^*_k$ we can determine the total 
forces acting on the particles and then, following 
Eq.~(\ref{particles-vel-update}) we obtain also the expected new 
velocities of the particles $\vec v^*_i$. The expected inner 
pressure, according to Eq.~(\ref{inner-pressure}), then reads:
\begin{equation}
  \ps = \frac{1}{\text{DV}}[\displaystyle\sum_{k=1}^{N_c} \vec
  F^*_k \cdot \vec l_k + \displaystyle\sum_{i=1}^{N} m_i \vec
  v^*_i \cdot \vec v^*_i].
  \label{expected-inner-pressure}
\end{equation}
Of course, there is no need to recalculate all the terms in 
Eq.~(\ref{expected-inner-pressure}) in order to update $\ps$. When 
the force at a single contact is changed it affects only three 
terms: one due to the force itself and two due to the velocities 
of the contacting particles. In order to save computational time, 
only the differences in these three terms have to be taken into 
account when $P^*_\text{in}$ is updated. Following
Eq.~(\ref{dilation-rate-update}), we obtain also the corresponding
value of the expected dilation rate:
\begin{equation}
 \ls = \lambda(t) +
 \frac{\ps-\Pext}{\Ml} \Delta t
  \label{expected-dilation-rate}
\end{equation}
This way, $\ls$ and $\ps$ are updated many times between two
consecutive time steps (in fact they are updated $N_I N_c$ times)
but in turn $\ls$ and $\ps$ are always consistent with the current
system of the contact forces. At the end of the iteration process
$\ps$ and $\ls$ provide not just an approximation of the new inner
pressure and new dilation rate but they are equal to 
$P_\text{in}(\tdt)$ and $\lambda(\tdt)$, respectively.

To complete the algorithm, we list here also the equations that 
are used for the force calculation of a single contact. The 
inequality (\ref{positive-gap}) is replaced by
\begin{equation}
  g + (\vrelfree + \ls \; \vec l) \cdot \n \Delta t > 0,
  \label{gap-check-new}
\end{equation}
i.e. there is no interaction between the two particles if the
inequality is satisfied. Otherwise we need a contact force. The
force, previously given by Eq.~(\ref{contact-force-gpos}), that is
required by a sticking contact is
\begin{equation}
  \vec R=\frac{-1}{\Delta t}{\bf M}
  (\frac{g^\text{pos}}{\Delta t}\n+\ls \; \vec l + \vrelfree).
  \label{contact-force-gpos-new}
\end{equation}
This force again has to be recalculated according to a sliding
contact if $\vec R$ in Eq.~(\ref{contact-force-gpos-new}) violates
the Coulomb condition:
\begin{equation}
  \vec R=\frac{-1}{\Delta t}{\bf M}
  (\frac{g^\text{pos}}{\Delta t}\n - \vrelt +\ls \; \vec l + \vrelfree),
  \label{contact-force-2-gpos-new}
\end{equation}
which replaces the original equation~(\ref{contact-force-2-gpos}).

Except the above changes, the CD algorithm remains the same. In
each time step the same iteration process is applied in order to
reach convergence of the contact forces. After the iteration
process we apply
Eqs.~(\ref{dilation-rate-update})-(\ref{particles-pos-update}) to
complete the time step.

The scheme of the solver for the modified version of CD can be
presented as
\begin{displaymath}
\hspace{-4cm}\left[
 \begin{array}{r}
  \hbox{$t=\tdt$ (time step)\hspace{5.95cm}} \\
  \hbox{Evaluating the gap $g$ for all contacts \hspace{3.65cm}} \\
  \left[
   \begin{array}{r}
    \hbox{$N_I=N_I+1$ (iteration)\hspace{5.3cm}} \\
    \left[
     \begin{array}{r}
      \hbox{$k=k+1$ (contact index)\hspace{4.65cm}} \\
      \hbox{Evaluating $\vrelfree_k$\hspace{5.8cm}}\\
      \hbox{Evaluating $\vec R_k$ (using Eqs.~(\ref{contact-force-gpos-new})
            and ~(\ref{contact-force-2-gpos-new}))\hspace{2.13cm}} \\
      \hbox{Evaluating $P^*_\text{in}$ (Eq.~(\ref{expected-inner-pressure})) then
      $\ls$ (Eq.(~\ref{expected-dilation-rate}))\hspace{1.83cm}} \\
     \end{array}
    \right.
    \\
    \hbox{Convergence test for contact forces\hspace{3.55cm}} \\
   \end{array}
  \right.
  \\
  \hbox{Time-stepping for the dilation rate and the system size (using
  Eqs.~(\ref{dilation-rate-update}) and~(\ref{system-update}))\hspace{-3.25cm}} \\
  \hbox{Time-stepping for velocities and positions of all
  particles (using Eqs.~(\ref{particles-vel-update})
  and~(\ref{particles-pos-update}))\hspace{-3.5cm}} \\
 \end{array}
\right.
\end{displaymath}
In the next section we will present some simulations with the
above method. We will test the algorithm and analyze the
properties of the resulting packings.

As an alternative to this fully implicit method we considered 
another possibility to discretize the
Eqs.~(\ref{dilation-rate})-({\ref{equation-motion-new-3}}) in the
spirit of the contact dynamics and, at the same time, impose the
constraint conditions on the new configuration. The main
difference is that the new value of the inner pressure
$P_{in}(\tdt)$ is determined based on the old velocities $\vec
v_i(t)$ and not on the new ones $\vec v_i(\tdt)$, while the
contribution of the forces are taken into account in the same way,
i.e. the new contact forces $\vec F_k(\tdt)$ are used in
Eq.~(\ref{inner-pressure}). Therefore this version of the method is 
only partially implicit, however, the constraint conditions and the 
force calculation 
[Eqs.~(\ref{gap-check-new})-(\ref{contact-force-2-gpos-new})] can
be applied in the same way. Only, the expected values $\ls$ and
$\ps$ has to be changed consistently with the new pressure
$P_{in}(\tdt)$:
\begin{equation}
  \ps= \frac{1}{\text{DV}}[\displaystyle\sum_{k=1}^{N_c} \vec F^*_k \cdot
  \vec l_k + \displaystyle\sum_{i=1}^{N} m_i \vec v_i(t) \cdot \vec v_i(t)]
  \label{expected-inner-pressure-notfullyimplicit}
\end{equation}
and then this modified $\ps$ is used to determine the expected
dilation rate $\ls$ with the help of
Eq.~(\ref{expected-dilation-rate}). Again here, $\ps$ and $\ls$
are calculated anew after each force update during the iteration
process and their last values equal the new pressure
$P_{in}(\tdt)$ and the new dilation rate $\lambda(\tdt)$.

We implemented and tested this second version of the method and
found that the constraint conditions are handled here also with
the same level of accuracy. Although the second method is perhaps
less transparent than the fully implicit version, for practical
applications it seems to be more useful. First, the second version
of the method is easier to implement into a program code, second, 
it turned out to be faster by $25 \%$ in our test simulations. The
improvement of the computational speed originates from the smaller
number of the operations. One does not have to handle the expected
particle velocities $\vec v^*_i$ and the recalculation of $\ps$ is
more simple as the change of a contact force $\vec F^*_k$ affects
only one term in the 
Eq.~(\ref{expected-inner-pressure-notfullyimplicit}). We note that 
here the distinction ``partially implicit'' and ``fully implicit'' 
refers only to the difference, whether the velocities are or are 
not included in the iteration process.

\begin{figure}[b]
\begin{center}
\includegraphics*[scale=0.45]{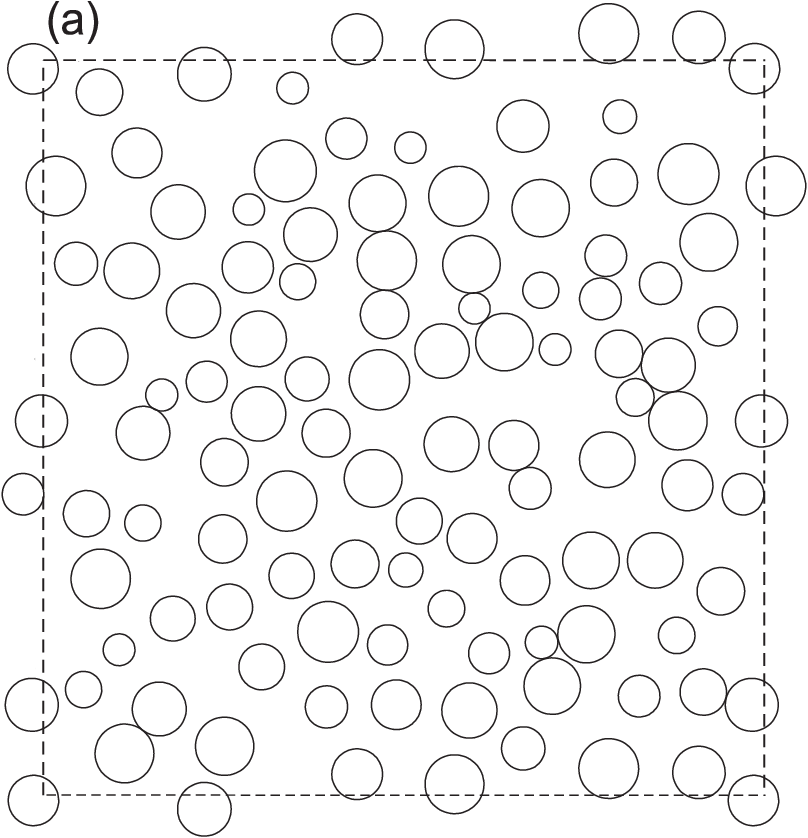}
\includegraphics*[scale=0.45]{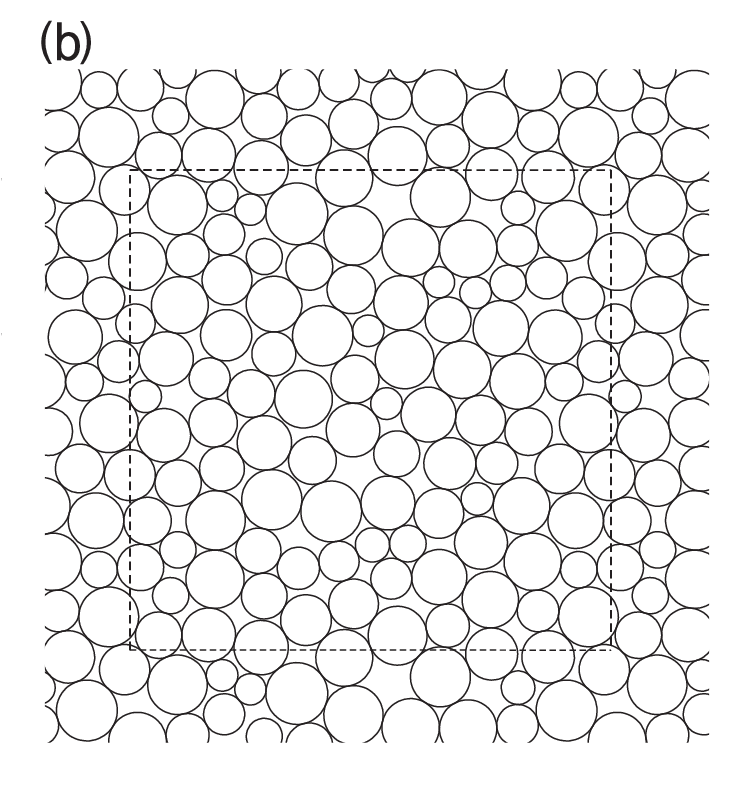}
\end{center}
\caption{Schematic picture of a two dimensional granular system 
controlled by a constant external pressure: (a) the initial gas 
state, and (b) the final homogeneous packing. The dashed lines 
mark periodic boundaries.}
\label{Fig-Schematic}
\end{figure}

\section{Numerical results}
\label{results}

We perform numerical simulations using the CD algorithm with the 
fully implicit constant pressure scheme of section~\ref{CD+Andersen}. 
This algorithm has been used to study mechanical properties of 
granular packings in response to local perturbations 
\cite{Shaebani07,Shaebani08pre}. Here, the main goal is to show 
that the algorithm works indeed in practical applications and to 
test the method from several aspects; We investigate how the 
simulation parameters influence the required CPU time and the 
accuracy of the simulation. Such parameters are the external 
pressure $\Pext$, the inertia parameter $\Ml$ and the 
computational parameters, like the number of iterations per time 
step $N_I$ and the length of the time step $\Delta t$. We also 
analyze the properties of the resulting packings.

Here, we report only simulations of two-dimensional systems of
disks, where the behavior is very similar to that we found for
spherical particles in three-dimensional systems. Length
parameters, the time and the two-dimensional mass density of the
particles are measured in arbitrary units of $l_0$, $t_0$ and
$\rho_0$, respectively. The samples are polydisperse and the disk
radii are distributed uniformly between 0.8 and 1.2, thus the
average grain radius is 1. The material of the grains has unit
density and the masses of the disks are proportional to their
areas. In this section we have one reference system that contains
$100$ disks. The interparticle friction coefficient is set to
$0.5$. The value of other parameters are: $N_I=100$, $\Delta
t=0.01$, $\Pext=1$ (this latter is expressed in units of 
$\rho_0 {l_0}\!^2 / {t_0}\!^2$) and the inertia $\Ml=100$ (in units 
of $\rho_0 {l_0}\!^2$). Throughout this section, we either use 
these reference parameters or the modified values will be given 
explicitly. Usually, we will vary only one parameter to check its 
effect while other parameters are kept fixed at their reference 
values.

\begin{figure}
\begin{center}
\includegraphics*[scale=1.0]{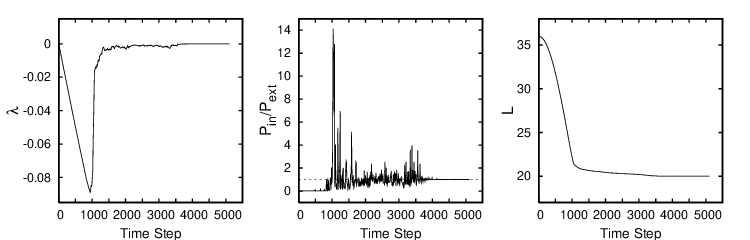}
\end{center}
\caption{Typical time evolution of the dilation rate $\lambda$
(left) the inner pressure $P_\text{in}$ (middle) and the system
size $L$ (right) during the compression of a 2D polydisperse
sample. The data shown here were recorded in the reference system
specified in the text.}
\label{Fig-TimeEvolutions}
\end{figure}

In each simulation, we start with a dilute sample of
nonoverlapping rigid disks randomly distributed in a two 
dimensional square-shaped cell [Fig.~\ref{Fig-Schematic}(a)]. No 
confining walls are used according to the boundary conditions 
specified in Sec.~\ref{CD+Andersen}. Gravity and the initial 
dilation rate are set to zero. Due to imposing a constant external 
pressure the dilute system starts shrinking. As the size of 
the cell decreases, particles collide, dissipate energy and after 
a while a contact force network is formed between touching 
particles in order to avoid interpenetrations. The contact forces 
build up the inner pressure $P_\text{in}$ which inhibits further 
contraction of the system. Finally, a static configuration is 
reached in which $P_\text{in}$ equals $\Pext$ and mechanical equilibrium is
provided for each particle [Fig.~\ref{Fig-Schematic}(b)]. 
Technically, we finish the simulation when the system is close 
enough to the equilibrium state: We apply a convergence 
threshold for the mean velocity $v_\text{mean}$ and 
mean acceleration $a_\text{mean}$ of the particles (which are
measured in units $l_0 / t_0$ and $l_0 / {t_0}\!^2$,
respectively). Only if both $v_\text{mean}$ and $a_\text{mean}$
become smaller than the threshold $10^{-10}$ we regard the system
as relaxed and stop the simulation.

The typical time evolution can be seen in
Fig.~\ref{Fig-TimeEvolutions} where we show the compaction process
in the case of the reference system.
Fig.~\ref{Fig-TimeEvolutions}(left) implies that the magnitude of
$\lambda$ grows linearly in the beginning when the inner pressure
is close to zero. The negative value of the dilation rate
indicates contraction which becomes slower after the particles
build up the inner pressure
[Fig.~\ref{Fig-TimeEvolutions}(middle)]. The fluctuations in
$P_\text{in}$ are due to collisions of the particles. In the final
stage of the compression $\lambda$ goes to zero, $P_\text{in}$
converges to the external pressure and the size of the system
reaches its final value [Fig.~\ref{Fig-TimeEvolutions}(right)].

\begin{figure}
\begin{center}
\includegraphics*[scale=1.0]{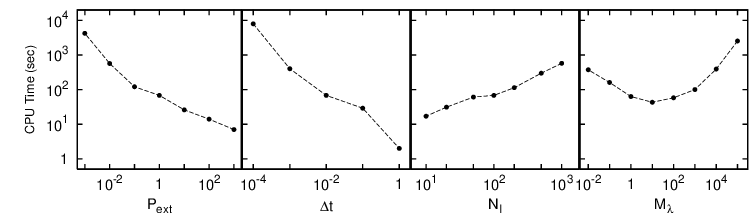}
\end{center}
\caption{CPU time versus the simulation parameters.}
\label{Fig-CPUtime}
\end{figure}

Next we investigate how the required CPU time of the simulation is
affected by the various parameters. All simulations are performed
with a processor Intel(R) Core(TM)2 CPU T7200 @ 2.00GHz and the
CPU time is measured in seconds. Figure~\ref{Fig-CPUtime} reveals
that the variation of $\Pext$, $\Delta t$, $N_I$ and $\Ml$ have
direct influence on the required CPU time. The final packing is
achieved with less computational expenses if larger $\Pext$,
larger $\Delta t$ or smaller $N_I$ is used. The role of system
inertia $\Ml$ is more complicated. $\Ml$ reflects the sensitivity of
the system to the pressure difference $P_\text{in} - \Pext$. If
the level of the sensitivity is too small or too large, the
simulation becomes inefficient. It is advantageous to choose the
inertia $\Ml$ near to its optimal value which depends on the
specific system (e.g. on the number and mass of the particles)
\cite{Kolb99}.

\begin{figure}[b]
\begin{center}
\includegraphics*[scale=1.0]{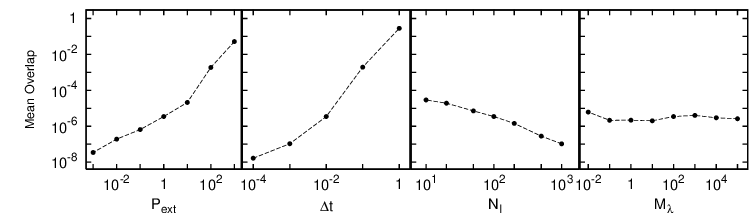}
\end{center}
\caption{Mean overlap in terms of the simulation parameters.}
\label{Fig-overlaps}
\end{figure}

\begin{figure}
\begin{center}
\includegraphics*[scale=0.49]{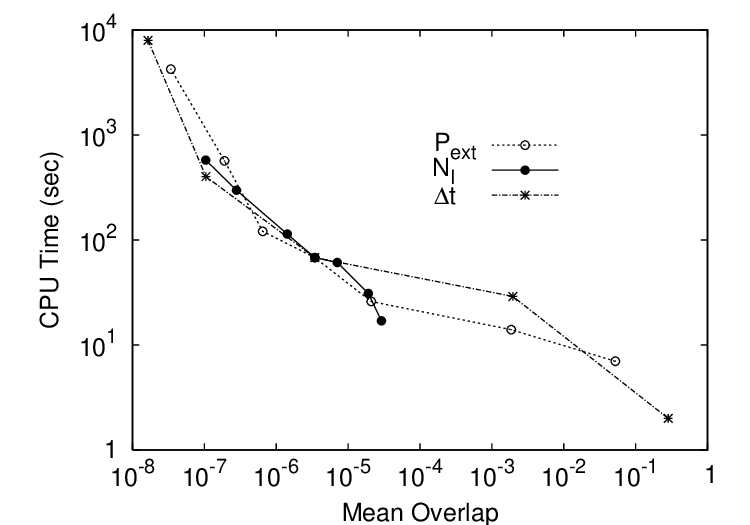}
\includegraphics*[scale=0.49]{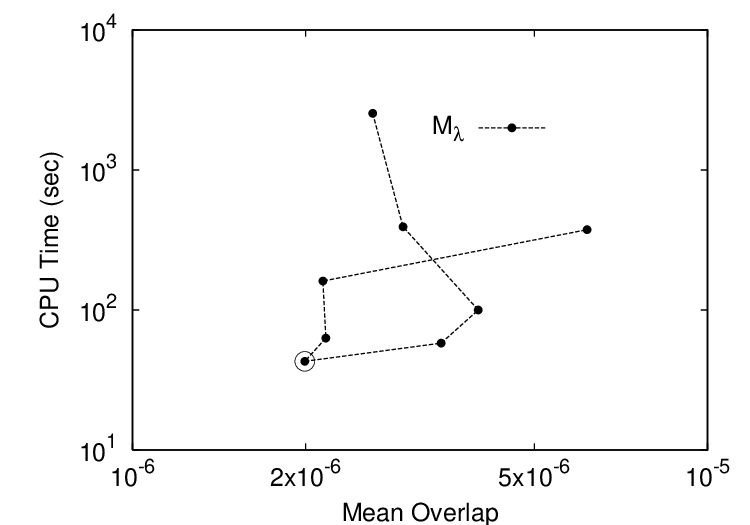}
\end{center}
\caption{CPU time in terms of the mean overlap. The different
curves are obtained by the variation of different parameters
according to Figs.~\ref{Fig-CPUtime} and \ref{Fig-overlaps}. These
parameters are the external pressure $\Pext$, the number of
iterations per time step $N_I$, the length of the time step
$\Delta t$ (left) and the inertia of the system $\Ml$ (right). The
open circle on the right indicates the most efficient simulation
we could achieve by controlling the inertia $\Ml$.}
\label{Fig-CPUtime-Overlap}
\end{figure}

\begin{figure}[b]
\begin{center}
\includegraphics*[scale=1.0]{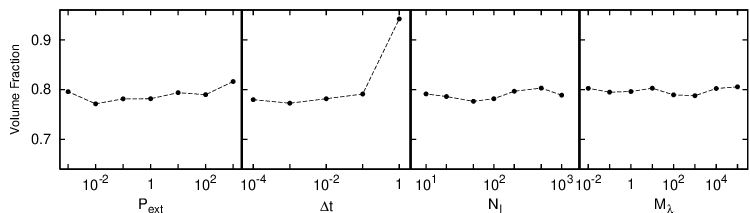}
\end{center}
\caption{Volume fraction versus the simulation parameters.}
\label{Fig-VolumeFraction}
\end{figure}

Regarding the efficiency of the computer simulation, not only the
computational expenses play an important role but the accuracy of
the simulation is also essential. Here we use the overlaps of the
particles as a measure of the inaccuracy of the simulation (see
Sec.~\ref{CD-Method}). In an ideal case there would be no overlaps
between perfectly rigid particles. Fig.~\ref{Fig-overlaps} shows
the mean overlaps measured in the final packings. It can be seen
for the parameters $\Pext$, $\Delta t$ and $N_I$ that the
reduction of the computational expenses at the same time leads
also to the reduction of the accuracy of the simulation. In
Fig.~\ref{Fig-CPUtime-Overlap}(left) we plot the CPU time versus
the mean overlap for these three parameters. The data points
collapse approximately on the same master curve which tells us
that the computational expenses are determined basically by the
desired accuracy; smaller errors require more computations. The
efficiency of the simulation is approximately independent of the
parameters $\Pext$, $\Delta t$ and $N_I$. The situation is,
however, different for the inertia of the system $\Ml$. First, it
has relatively small effect on the accuracy of the simulation
(Fig.~\ref{Fig-overlaps}). Variation of $\Ml$ by $7$ orders of
magnitude could hardly change the mean overlap of the particles.
Second, $\Ml$ affects strongly the efficiency of the simulation
which is shown clearly by Fig.~\ref{Fig-CPUtime-Overlap}(right).
If $\Ml$ is varied then larger computational expense is not
necessarily accompanied with smaller errors. In fact, in the whole
range of $\Ml$ studied here, the fastest simulation turned out to be
the most accurate one [open circle in
Fig.~\ref{Fig-CPUtime-Overlap}(right)].

\begin{figure}
\begin{center}
\includegraphics*[scale=0.37]{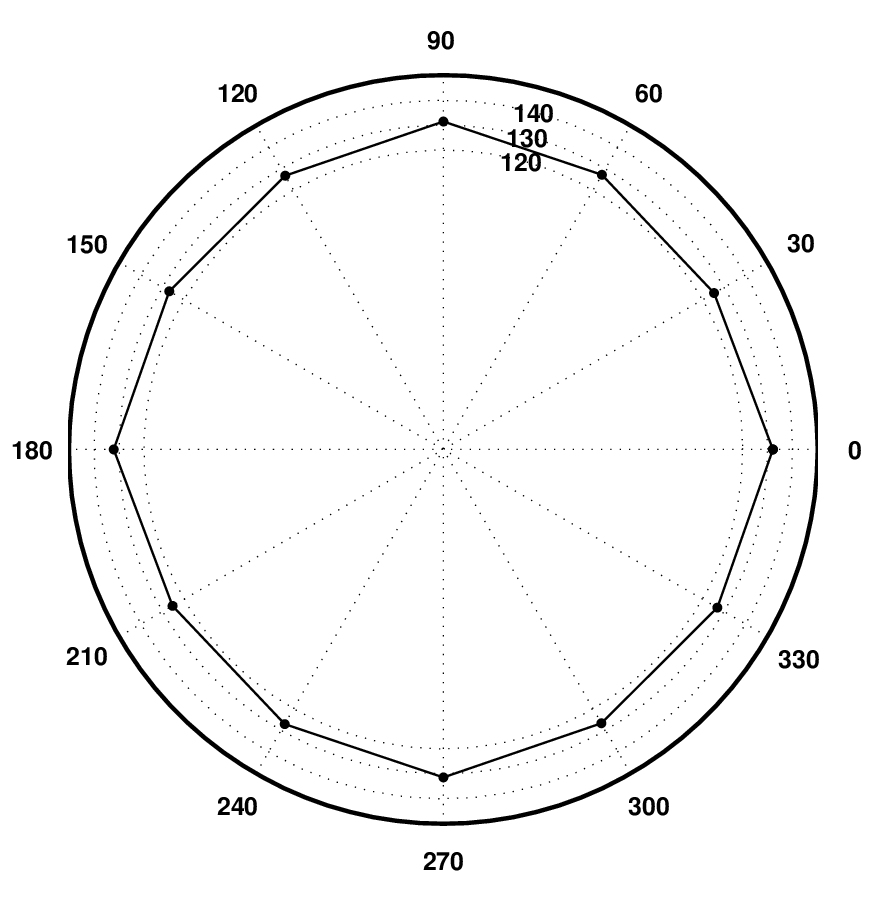}
\includegraphics*[scale=0.55]{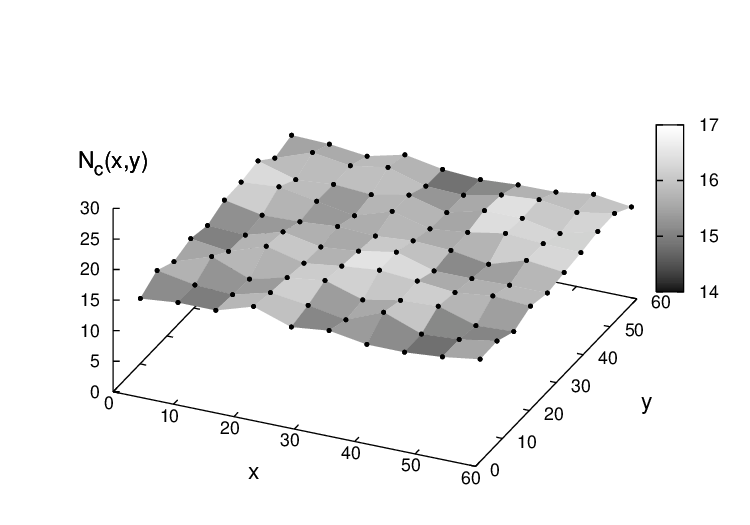}
\end{center}
\caption{(left) The angular distribution of the contacts. The
number of contacts is plotted in terms of the direction of their
normal vectors. (right) Spatial distribution of the contacts. The
system contains $N = 1000$ disks in both figures.}
\label{Fig-ContactDensity}
\end{figure}

Next we turn to the question whether the parameters of the
simulation used in the compaction process influence the physical
properties of the final packing. There are many ways to
characterize static packings of disks. Here we test only one
quantity, the frequently used volume fraction. The volume fraction
gives the ratio between the total volume (total area in 2D) of
grains and the volume (area) of the system.
Fig.~\ref{Fig-VolumeFraction} shows the volume fraction of the
same packings that were studied already in
Fig.~\ref{Fig-overlaps}. It can be seen that the volume fraction
remains approximately unchanged under the variation of the four
parameters $\Pext$, $N_I$, $\Delta t$ and $\Ml$. This is except for
one data point for large time step $\Delta t$, where the
simulation is very inaccurate. The corresponding mean overlap
(Fig.~\ref{Fig-overlaps}) is comparable to the typical size of the
particles which is the reason why the volume fraction appears to
be much larger.

Finally, we investigate the inner structure of the resulting
random packings. For that, we study larger samples with $N=1000$
particles, otherwise, the default parameters are used during the
compaction. To suppress random fluctuations, we produce $5$
different systems and all quantities reported hereafter represent
average values over these systems. In
Fig.~\ref{Fig-ContactDensity}(left) we study the angular
distribution of the contact normals and find that it is very close
to uniform. However, there is a small but definite deviation
(around $3 \%$): the density of the contact normals are slightly
larger parallel to the periodic boundaries. In this sense the
packing is not completely isotropic. Although the effect is very
small, the orientation of the boundaries can be observed also in
such local quantities like the direction of the contacts.

In connection to the question of the isotropy we checked also the
global stress tensor $\sigma$. In the original frame $\sigma$
reads
\begin{equation}
 \sigma = \left(\begin{array}{cc} 1.00909 & -0.01334 \\
          -0.01332 & 0.99091 \end{array} \right).
  \label{stress-value}
\end{equation}
This stress is isotropic with good approximation. The diagonal
entries are close to $1$ which equals $\Pext$ while the
off-diagonal elements are approximately zero. Compared to the unit
matrix, the elements deviate around $1 \%$ of $\Pext$.

The final packings are expected to be homogeneous as all points of
the space are handled equivalently by the compaction method. Apart 
from random fluctuations we do not observe any inhomogeneity in 
our test systems. As an example, we show the spatial distribution 
of the contacts in Fig.~\ref{Fig-ContactDensity}(right), where the
density is approximately constant.

\section{Concluding remarks}
\label{conclusions}

In this work we have proposed and tested a simulation method to
produce homogeneous random packings in the absence of confining
walls. We combined the contact dynamics algorithm 
with a modified version of the Andersen method to perform constant 
pressure simulations of granular systems. Our main concern
was to discuss how constraint conditions can be applied to
determine the interaction between the particles in an
Andersen-type of dynamics. We have presented the results of some
numerical tests and discussed the effect of the main parameters on
the efficiency of the simulations and on the physical properties
of the final packings.

We restricted our study to the simple case where we allow only
spherical strain of the system in order to achieve the desired
pressure. However, the method can be generalized to apply other
type of constraints to the stress tensor and, consequently, to
allow more general strain deformations where shape as well as size 
of the simulation cell can be varied \cite{Parrinello80,Allen96}.

\section*{Acknowledgements}
We acknowledge support by grants No.~OTKA T049403, No.~OTKA
PD073172 and by the Bolyai Janos Scholarship of the Hungarian 
Academy of Sciences.

\end{document}